# Reduced Coercive Field in Epitaxial Thin Film of Ferroelectric Wurtzite Al$_{0.7}$Sc$_{0.3}$N

*Keisuke Yazawa[1,2]\*, Daniel Drury[1,2], Andriy Zakutayev[1], and Geoff L. Brennecka[2]\**

1. Materials Science Center, National Renewable Energy Laboratory, Golden, Colorado 80401, United States

2. Department of Metallurgical and Materials Engineering, Colorado School of Mines, Golden, Colorado 80401, United States

**Corresponding Authors**

*E-mail: yazawa@mines.edu, geoff.brennecka@mines.edu

**ABSTRACT**

The first epitaxial ferroelectric wurtzite film with clear polarization-electric field hysteresis behavior is presented. The coercive field of this epitaxial Al$_{0.7}$Sc$_{0.3}$N film on W/*c*-sapphire substrate is 0.4 ± 0.3 MV cm$^{-1}$ (8 %) smaller than that of a conventional fiber textured film on a Pt/TiO$_x$/SiO$_2$/Si substrate, attributed to the 0.01 ± 0.007 Å smaller *c*-axis lattice parameter in the epitaxial film. The strain and decrease of the coercive field most likely originate from epitaxial strain rather than the mismatch in thermal coefficient of expansion. These results provide an insight for further coercive field reduction of novel wurtzite ferroelectrics using epitaxial mismatch strain.

Nitride wurtzite materials form the basis of a wide range of semiconductor, piezoelectric, and optoelectronics devices.[1–3] Recent reports of ferroelectricity in $Al_{1-x}Sc_xN$ films[4,5] promise to expand the application space even further, but broad adoption in low power devices will likely require significant reduction of the coercive fields (~3-5 MV cm$^{-1}$). Increasing Sc content in $Al_{1-x}Sc_xN$ reduces the coercive field because of the corresponding reduction in the *c/a* ratio.[4,5] However, this method is limited in effect because increased Sc content is also associated with degraded structural quality due to the strong thermodynamic driving force for phase separation[6] as well as increased leakage, in part because of the much smaller band gap of ScN.[7] Ideally, one would like to significantly reduce coercive field without degrading band gap or, more directly, breakdown field.

A complementary approach to alloying is strain engineering, as the coercive field of wurtzite ferroelectrics is relatively sensitive to mechanical stresses,[4,8] similar to conventional perovskite ferroelectrics, attributed to the electromechanical thermodynamic coupling.[9,10] It is known that sputtering process parameters influence residual stress, and both substrate stacks and deposition parameters are commonly used to engineer stress states in AlN and related films.[11,12] Strain engineering via epitaxial growth has become commonplace in many systems[13,14] but has not previously been applied to tune the coercive field in the $Al_{1-x}Sc_xN$ system. An additional potential benefit of epitaxial confinement is the stabilization of the wurtzite phase in high Sc content $Al_{1-x}Sc_xN$, analogous to realizing a metastable perovskite phase via epitaxial growth.[15–17] Several prior studies have grown epitaxial wurtzite films on sapphire substrates, and the mechanism and epitaxial relationship have been thoroughly investigated.[18] Although $Al_{1-x}Sc_xN$ epitaxial films on sapphire substrates have also been reported[19–23] no ferroelectric response was reported for those films.

To perform ferroelectric switching, fabricating a parallel capacitor device is favorable. The ideal bottom electrode between the film and the sapphire substrate should be a material with high chemical stability for sharp interfaces grown epitaxially on the sapphire substrate. BCC tungsten has been grown epitaxially on sapphire substrates: W (111) out of plane orientation on c-sapphire via chemical vapor deposition[24] and W (100) on *r*-sapphire via electron beam evaporation and chemical vapor deposition method.[24,25] Also, the formation energy of hexagonal $WN_2$ (-50 kJ mol$^{-1}$) or WN (-25 kJ mol$^{-1}$) is less negative than that of AlN (-145 kJ mol$^{-1}$)[26], so that interdiffusion via the reactions $W + 2AlN \rightarrow WN_2 + 2Al$ or $W + AlN \rightarrow WN + Al$ is thermodynamically unfavorable, and a sharp interface can be maintained.

In this study, the first epitaxial ferroelectric $Al_{0.7}Sc_{0.3}N$ film on W/c-sapphire is presented. The in-plane alignment for the epitaxial film is implied as [110] AlScN//[001] W and [1-10] W//[110] sapphire. The epitaxial film exhibits considerable decrease of the out-of-plane lattice parameter $c$ (~0.01 ± 0.007 Å) as well as coercive field (~400 ± 300 kV cm$^{-1}$) when compared to a conventional fiber-textured film on a platinized silicon substrate. The origin of the lattice parameter and coercive field decrease is likely from epitaxial strain rather than mismatch in thermal coefficient of expansion. This study therefore suggests that epitaxy can be an additional powerful tool for engineering the coercive field of uniaxial ferroelectrics such as $Al_{1-x}Sc_xN$.

Epitaxial W and $Al_{0.7}Sc_{0.3}N$ were deposited via (reactive) RF magnetron sputtering using baseline growth conditions as follows: 5 mtorr of Ar, target-substrate distance of 129 mm, and target power of 7.7 W/cm$^2$ on a 3" diameter elemental W target, and 2 mtorr of Ar/N$_2$ (15/5 sccm flow), target-substrate distance of 165 mm, and target power of 6.6 W/cm$^2$ on a 3" diameter

Al$_{0.7}$Sc$_{0.3}$ alloy target (Stanford Advanced Materials) for Al$_{0.7}$Sc$_{0.3}$N growth. Slight changes in Al$_{0.7}$Sc$_{0.3}$N deposition condition (Ar/N$_2$ = 14/6 sccm, 183 mm target-substrate distance, 1.5 mtorr deposition pressure) were used to induce variations in structure and properties of the films, resulting in thickness variation in the 200 – 270 nm range. The substrate temperature was 400 °C and the substrate was rotated for both depositions. For Al$_{0.7}$Sc$_{0.3}$N growth, the base pressure, partial oxygen and water vapor pressure at 400 °C are $< 5 \times 10^{-7}$ torr, $P_{O2} < 2 \times 10^{-8}$ torr and $P_{H2O} < 3 \times 10^{-7}$ torr, respectively. The W/$c$-sapphire substrate as well as a conventional Pt/TiO$_x$/SiO$_2$/Si substrate[27] were mounted side-by-side for direct comparison of the subsequent films. The crystal structure of the films is investigated using X-ray diffraction (Panalytical Empyrean). Top Pt contacts 110 µm in diameter were deposited on the Al$_{0.7}$Sc$_{0.3}$N films via DC sputtering through a shadow mask for ferroelectric (*P-E*) measurements using a Precision Multiferroic system from Radiant Technologies.

The XRD data for crystal structure analysis is shown in Figure 1. Typical out-of-plane $\theta$-$2\theta$ profiles for the films on W/$c$-sapphire and Pt/TiO$_x$/SiO$_2$/Si substrates indicate the pure wurtzite Al$_{0.7}$Sc$_{0.3}$N phase with the out-of-plane (001) texture (Figure 1(a)). In addition, W (110) is also preferably oriented on $c$-sapphire unlike a previous study showing (111)W/c-sapphire.[24] This orientation variation likely originates from the (110) and (111) surface energy competing on the c-sapphire plane; the triangular matching of W (111) to sapphire (001) lowers the W (111) surface energy relative to the W (110) surface energy, which is the lowest-energy surface for a BCC metal in vacuum (Figure S1). This is directly analogous to epitaxial BCC Ta / c-sapphire reported previously.[28] In this current study, W (110) is the exclusively dominant orientation.

The pole figure of the off-axis (103) plane of the Al$_{0.7}$Sc$_{0.3}$N film on W/$c$-sapphire exhibits 6-fold symmetry, while the Al$_{0.7}$Sc$_{0.3}$N film on Pt/TiO$_x$/SiO$_2$/Si exhibits random in-plane orientation (Figure 1(b)). Thus, the Al$_{0.7}$Sc$_{0.3}$N film grows epitaxially on the W/*c*-sapphire substrate, whereas the film on the Pt/TiO$_x$/SiO$_2$/Si substrate possesses a fiber texture. The epitaxial relationship among Al$_{0.7}$Sc$_{0.3}$N, W and sapphire is that W [1-10] aligns to *c*-sapphire [110], and Al$_{0.7}$Sc$_{0.3}$N is 30° rotated with regard to *c*-sapphire based on the pole figures of *c*-sapphire, W and Al$_{0.7}$Sc$_{0.3}$N showing 6-fold symmetry and appear at the same phi angle (Figure S2). Note that W has three variants. Although this c-sapphire - wurtzite 30° rotation was widely reported and rigorously investigated in direct epitaxy on c-sapphire,[18] the mechanism in our case is different from the earlier study due to the existence of the W layer between them. The prior study explained this orientation preference in terms of the Al and Sc ions being comfortably accommodated in an oxygen tetrahedral coordination site at the sapphire surface,[18] but the W (110) surface has no such oxygen sublattice site. Given the bulk lattice parameters of these materials, the Al$_{0.7}$Sc$_{0.3}$N *a*-axis most likely aligns to the in-plane W *c*-axis, so that Al$_{0.7}$Sc$_{0.3}$N rotates 30° against *c*-sapphire (Figure 1(c)).

Reciprocal space maps used for detailed crystal structure analysis are shown in Figure 1(d). Upper figures are the maps on a plane in reciprocal space including sapphire (006) and (110) reciprocal lattice vectors (phi = 0) whereas bottom figures are on a plane possessing (006) and (210) reciprocal lattice vectors (phi = 30°). From the positions of the W (220), (400) and (222) reciprocal lattice vectors, the W lattice structure is: $a = b = 3.169$ Å, $c = 3.157$ Å and $\gamma = 90.0°$. Thus, the W is tetragonally distorted ($c/a = 0.996$), likely caused by partial epitaxial confinement at the semi-coherent interface as the W [1-10] is not fully constrained to the sapphire [110]. This W strain is unlikely originated from the thermal expansion coefficient mismatch strain since the W thermal expansion coefficient ($5 \times 10^{-6}$ K$^{-1}$)[29] is smaller than that of a-axis sapphire ($7.3 \times 10^{-6}$ K$^{-1}$).[30] The positions of the Al$_{0.7}$Sc$_{0.3}$N film (004) and (114) reciprocal space vectors show the lattice parameters to be $a = 3.252$ and $c = 4.979$ ($c/a = 1.531$). These values are comparable to other reported $Al_{1-x}Sc_xN$ thin films[3,5] and show that the $Al_{0.7}Sc_{0.3}N$ [110] is also not fully constrained to the W [001].

Figure 2(a) shows *P-E* hysteresis loops for Pt/$Al_{0.7}Sc_{0.3}$N/W/*c*-sapphire and Pt/$Al_{0.7}Sc_{0.3}$N/Pt/$TiO_x$/$SiO_2$/Si capacitors measured using an excitation electric field of 10 kHz. The nested loops were measured for gradually-reduced maximum electric fields. Clear well-saturated ferroelectric hysteresis loops are observed for both the epitaxial and fiber-textured films, proving the ferroelectric switching occurs for both the epitaxial and fiber films. At the highest applied electric field (~6 MV cm$^{-1}$), polarization swelling (~10 µC cm$^{-2}$) for both the films is observed, which is most likely attributed to leakage current contribution.[31]

Figure 2(b) shows the remanent polarization $(+P_r-(-P_r)/2)$ as a function of the applied electric fields for the two films. The polarization is well saturated above the respective coercive fields, reaching values of ~120 µC cm$^{-2}$, consistent with literature values for films of similar composition.[4,5] It is noteworthy that the coercive field of the epitaxial film is considerably smaller than that of the fiber-textured film by ~400 kV cm$^{-1}$ (8 %), which is important for practical applications. However, the slopes of the *P-E* loops during switching are noticeably steeper for the fiber-textured film on Pt/$TiO_x$/$SiO_2$/Si than those of the epitaxial film on W/*c*-sapphire. To understand the origin of this difference, further investigation into switching dynamics is required.

To correlate the crystal structure/texture and ferroelectric coercive field, the coercive field $(+E_c-(-E_c)/2)$ as a function of the out of plane lattice spacing is illustrated in Figure 3(a). The coercive field has a strong linear correlation to the lattice parameter, i.e., a film with a smaller *c*-axis has a smaller coercive field with the ~400 kV cm$^{-1}$/ 0.01 Å slope. This can directly relate to the reported stress effect on the coercive field in the system[4] under the assumption that the strain is in the elastic region, namely an in-plane compressive stress increases the coercive field. The lattice spacing variation in our experiment likely originated from stresses coming from process parameters: thickness and the sputtering conditions described in methods section.[11] The $c$-axis strain range estimated from the lattice spacing obtained by XRD is less than 0.8 %, corresponding to $< \pm 1.2$ GPa in biaxial in-plane stress, which is in a reasonable range of the process induced residual stress in magnetron sputtering.[11]

The slope of the correlation between the lattice parameter and coercive field (~400 kV cm$^{-1}$/ 0.01 Å) is, however, smaller than the estimated slope based on the reported linear relationship between in-plane stress and coercive field (~800 kV cm$^{-1}$/GPa)[4] and the compliance tensor. With the in-plane stress, the out-of-plane strain $\varepsilon_{33}$ is calculated assuming the elastic regime, thereby using Hooke's low $\varepsilon_{33} = s_{33ij}\sigma_{ij}$ and $\sigma_{11} = \sigma_{22}$, $\sigma_{33} = \sigma_{23} = \sigma_{13} = \sigma_{12} = 0$ with a reported compliance tensor,[32] and the slope of (002) spacing change – coercive field line turns to be ~1000 kV cm$^{-1}$/ 0.01 Å. This more than factor of two difference in the slope of the lattice parameter-coercive field relationship can be caused by the gap between the macroscopic stress and crystal strain; the macroscopic stress can also contribute to creep and grain boundary sliding in addition to the lattice strain. In other words, our data provides the pure crystal strain – coercive field relationship excluding the other extrinsic components seen in a macroscopic stress measurement.

Most remarkably, the epitaxial films on W/$c$-sapphire tend to have smaller out-of-plane lattice parameters $c$ and smaller coercive fields. The direct comparison between films by substrates is shown in Figure 3(b). The bars represent the average value, and the error bars show the standard deviation for each. The epitaxial films exhibit a significant decrease in both the lattice spacing (by ~0.01 ± 0.007 Å) and coercive field (by ~400 ± 300 kV cm$^{-1}$). While the possible sources causing the difference depending on the substrates can be the partial epitaxial confinement strain and/or thermal expansion coefficient (CTE) mismatch strain, the epitaxial strain is more likely the origin of the in-plane tensile stress and resulting decrease of the coercive field compared to conventional fiber textured films on a Pt/TiO$_x$/SiO$_2$/Si substrate. The thermal expansion coefficients are $3.6 \times 10^{-6}$ K$^{-1}$ for Si, $7.3 \times 10^{-6}$ K$^{-1}$ for in-plane $c$-sapphire[30] and $4 \sim 5 \times 10^{-6}$ K$^{-1}$ for in-plane Al$_{1-x}$Sc$_x$N,[33] so that a $c$-sapphire substrate ought to give more compressive in-plane strain from CTE mismatch alone to the Al$_{0.7}$Sc$_{0.3}$N film than a Si substrate does, which opposes what has been observed here. Although further studies such as investigating the macroscopic mechanical stress-strain relationship and/or growing films on a range of substrate types need to be conducted to quantify the epitaxial strain effect, this insight is the first step to exploring the tunability of coercive field via epitaxy in ferroelectric wurtzite materials.

In summary, the first ferroelectric epitaxial (001) Al$_{0.7}$Sc$_{0.3}$N film on a (110) W/$c$-sapphire substrate was achieved using RF magnetron sputtering. The epitaxial relationship was implied as [110] AlScN//[001] W and [1-10] W//[110] sapphire. The epitaxial film exhibited a well saturated *P-E* hysteresis loop comparable to conventional fiber textured film. The epitaxial films have 0.01 ± 0.007 Å smaller out-of-plane lattice parameter and 400 ± 300 kV cm$^{-1}$ smaller ferroelectric coercive field on average compared to fiber textured Al$_{0.7}$Sc$_{0.3}$N films on a Pt/TiO$_x$/SiO$_2$/Si substrates. This result shows not only the applicability of the epitaxial film for decreasing the switching voltage in ferroelectric devices, but also new guideline to tailor the coercive field in the novel ferroelectric wurtzite film.

See supplementary materials for further detail and data for epitaxial relationship.

This work was co-authored by Colorado School of Mines and the National Renewable Energy Laboratory, operated by the Alliance for Sustainable Energy, LLC, for the U.S. Department of Energy (DOE) under Contract No. DE-AC36-08GO28308. Funding was provided by the DARPA Tunable Ferroelectric Nitrides (TUFEN) program (DARPA-PA-19-04-03) as a part of Development and Exploration of FerroElectric Nitride Semiconductors (DEFENSE) project (structural and electrical characterization), and by Office of Science (SC), Office of Basic Energy Sciences (BES) as part of the Early Career Award "Kinetic Synthesis of Metastable Nitrides" (material synthesis). The authors also express their appreciation to John Hayden and Prof. Jon-Paul Maria of the Pennsylvania State University for technical advice regarding W sputtering conditions and to Dr. Wanlin Zhu and Prof. Susan Trolier-McKinstry for providing Pt/TiO$_x$/SiO$_2$/Si Pt substrates. We also thank Dr Kevin Talley and Dr Jeff Alleman for helping sputtering system setup at NREL. The data affiliated with this study are available from the corresponding author upon reasonable request. The views expressed in the article do not necessarily represent the views of the DOE or the U.S. Government.

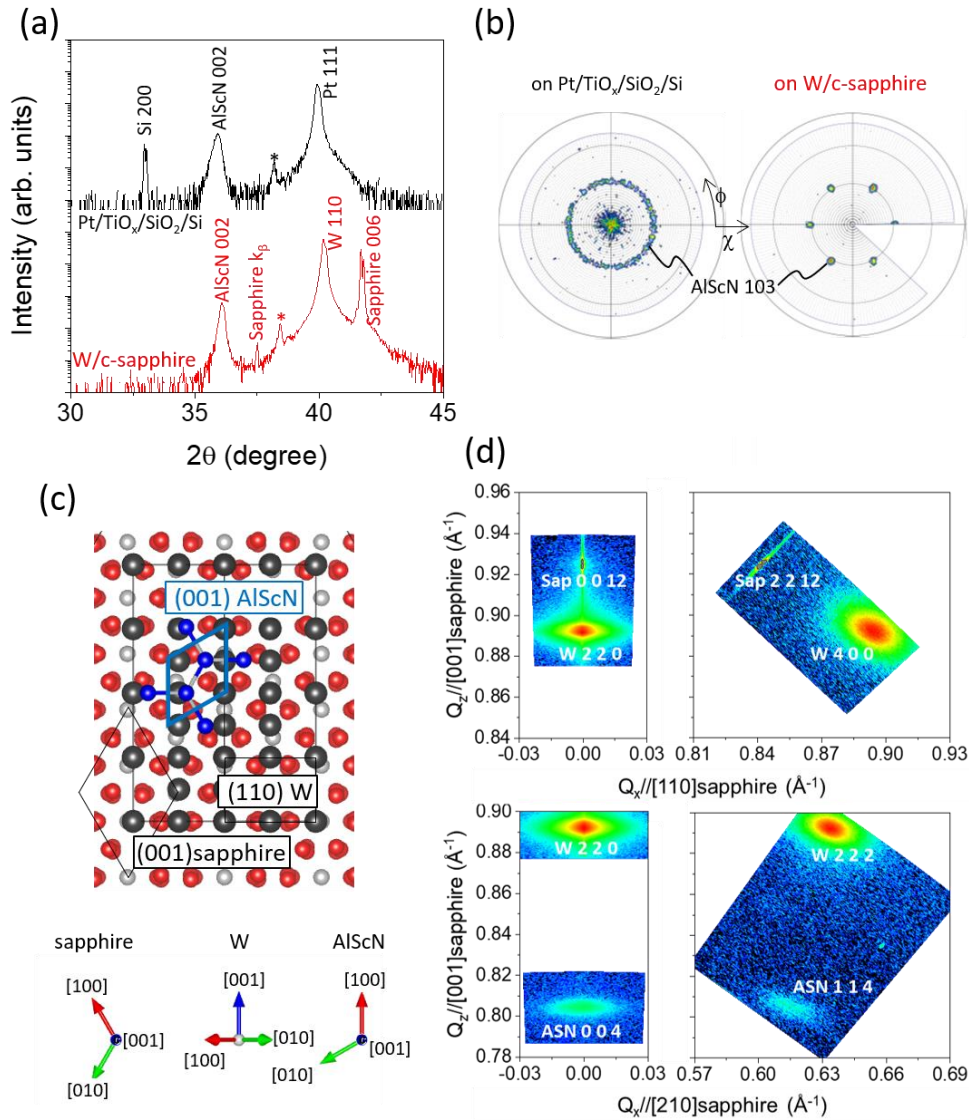

FIG. 1 Crystallographic analysis of $Al_{0.7}Sc_{0.3}N$ thin films. (a) XRD $\theta$-$2\theta$ patterns and (b) pole figures for films on W/$c$-sapphire and Pt/TiO$_x$/SiO$_2$/Si substrates. (c) Schematic of epitaxial relation of $Al_{0.7}Sc_{0.3}N$/W/$c$-sapphire using VESTA.[34] (d) Reciprocal space maps of $Al_{0.7}Sc_{0.3}N$ film on W/$c$-sapphire. Upper figures show a plane in the reciprocal space including [001] and [110] of sapphire substrate while bottom figures represent a plane including [001] and [210] of sapphire substrate.

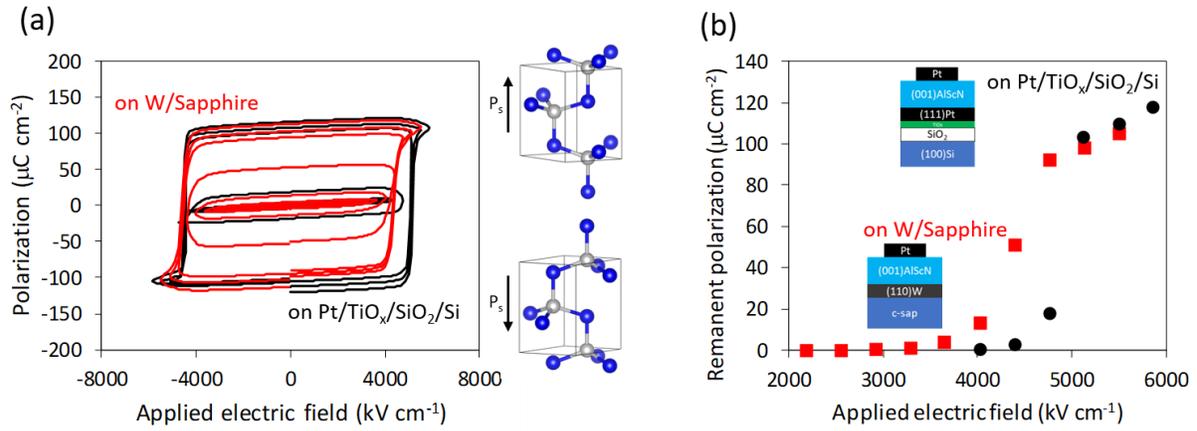

FIG. 2 Ferroelectric properties for $Al_{0.7}Sc_{0.3}N$ thin films. (a) Hysteresis loops of epitaxial film on W/$c$-sapphire and fiber film on Pt/TiO$_x$/SiO$_2$/Si substrates. The excitation frequency was 10 kHz. (b) Polarization saturation property of epitaxial film on W/c-sapphire and fiber film on Pt/TiO$_x$/SiO$_2$/Si substrates.

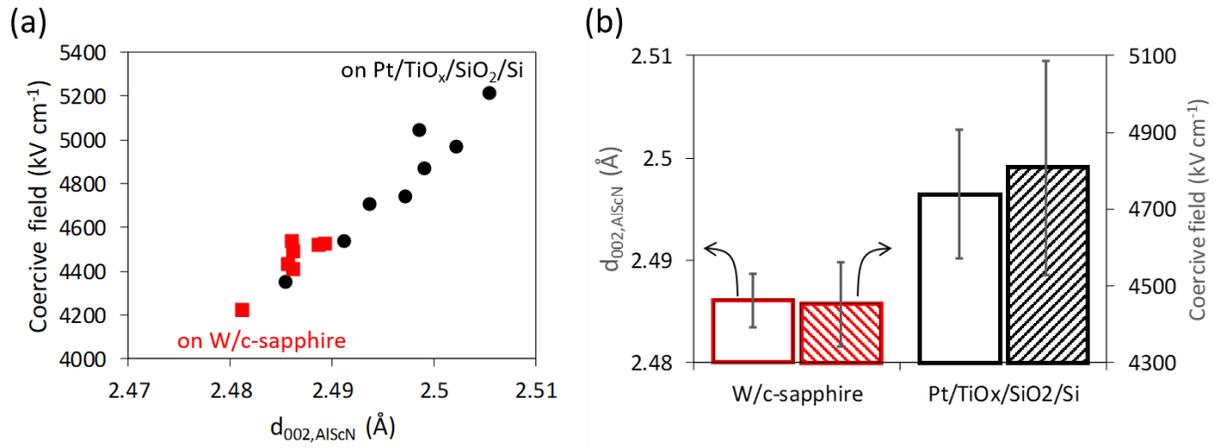

FIG. 3 Coercive field and $d_{002}$ lattice space comparison between epitaxial $Al_{0.7}Sc_{0.3}N$ film on W/*c*-sapphire and fiber film on $Pt/TiO_x/SiO_2/Si$ substrates. (a) Correlation between coercive field and $d_{002}$ in both epitaxial and fiber films. (b) Statistical comparison in $Al_{0.7}Sc_{0.3}N$ $d_{002}$ lattice spacing and coercive field between epitaxial and fiber films. Open bars represent lattice spacing while shadowed bars show coercive field.

**Supplemental Materials**

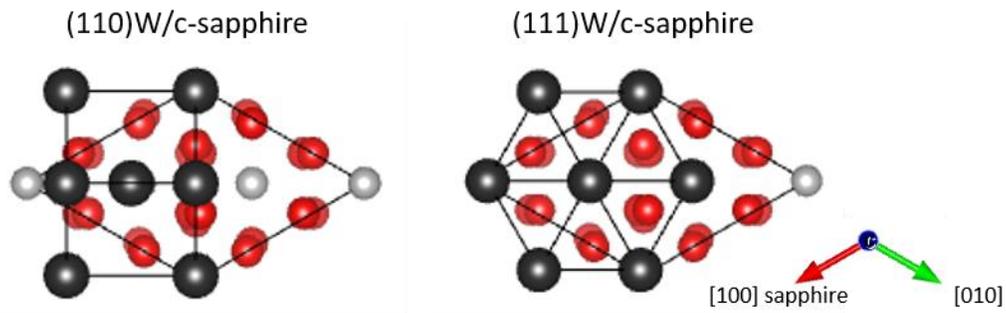

FIG. S1 Top views of schematic (110)W stack and (111)W stack on *c*-sapphire. Black, red and grey atoms represent W, O and Al, respectively. Energetically preferable stack can be in competition: intrinsic lowest surface energy plane (110) vs better lattice matching between (111) W and (001) sapphire.

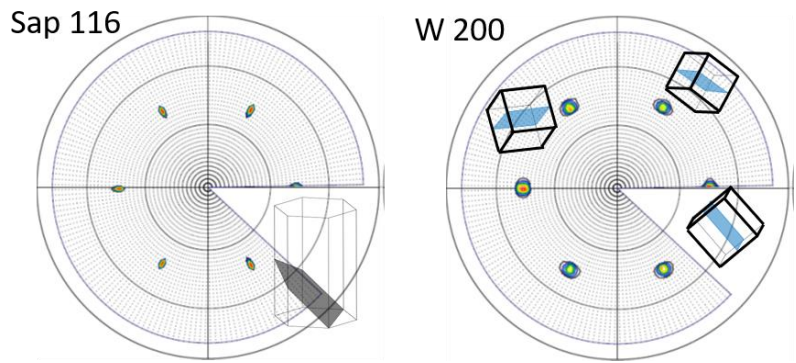

FIG. S2 Pole figures for 116 pole of *c*-sapphire and 200 pole of (110) W. There are three W in-plane orientation variants based on the six spots originated from 2-fold 200 pole.